\documentclass[letterpaper,12pt]{article}
\usepackage{color,amssymb,enumerate,float}

\usepackage{ifpdf}
\ifpdf
	\usepackage[pdftex]{graphicx}
	\usepackage[pdftex,unicode,implicit]{hyperref}

	\hypersetup{
  	pdftitle     = {}, 
  	pdfkeywords  = {},
  	pdfauthor    = {},
  	pdfcreator   = {pdf\LaTeXe\ with package \flqq hyperref\frqq},
  	pdfproducer  = {pdf\LaTeXe\ with package \flqq hyperref\frqq},
  	pdfpagemode  = UseNone,  
  	pdffitwindow = true,  
  	unicode      = true,
  	plainpages   = true,
  	colorlinks   = true,  
  	citecolor    = black,  
  	urlcolor     = blue, 
  	linkcolor    = black
	}

\else

  \usepackage[dvips]{graphicx}

\fi 

\makeatletter
\@addtoreset{equation}{section}
\makeatother

\begin{document}
	
\thispagestyle{empty}

\begin{center}
{\bf \LARGE Dynamics of the anisotropic conformal Ho\v{r}ava theory versus its kinetic-conformal formulation}
\vspace*{15mm}

{\large Jorge Bellor\'{\i}n}$^{1}$
{\large and Byron Droguett}$^{2}$
\vspace{3ex}

{\it Department of Physics, Universidad de Antofagasta, 1240000 Antofagasta, Chile.}
\vspace{3ex}

$^1${\tt jbellori@gmail.com,} \hspace{1em}
$^2${\tt byron.droguett@ua.cl}

\vspace*{15mm}
{\bf Abstract}
\begin{quotation}{\small  
	We contrast the dynamics of the Ho\v{r}ava theory with anisotropic Weyl symmetry with the case when this symmetry is explicitly broken, which is called the kinetic-conformal Ho\v{r}ava theory.
	To do so we perform the canonical formulation of the anisotropic conformal theory at the classical level with a general conformal potential. Both theories have the generator of the anisotropic Weyl transformations as a constraint but it changes from first to second-class when the Weyl symmetry is broken. The FDiff-covariant vector $a_i = \partial_i \ln N$ plays the role of gauge connection for the anisotropic Weyl transformations. A Lagrange multiplier plays also the role of gauge connection, being the time component. The anisotropic conformal theory propagates the same number of degrees of freedom of the kinetic-conformal theory, which in turn are the same of General Relativity. This is due to exchange of a second-class constraint in the kinetic-conformal case by a gauge symmetry in the anisotropic conformal case. An exception occurs if the conformal potential does not depend on the lapse function $N$, as is the case of the so called $(\mbox{Cotton})^2$ potential, in which case one of the physical modes becomes odd. We develop in detail two explicit anisotropic conformal models. One of them depends on $N$ whereas the other one is the $(\mbox{Cotton})^2$ model. We also study conformally flat solutions in the anisotropic conformal and the kinetic-conformal theories, defining as conformally flat the spatial metric, but leaving for $N$ a form different to the one dictated by the anisotropic Weyl transformations. We find that in both theories these configurations have vanishing canonical momentum and they are critical points of the potential. In the kinetic-conformal theory we find explicitly an exact, nontrivial, conformally flat solution. 
}
   \end{quotation}

\end{center}

\thispagestyle{empty}

\newpage
\section{Introduction}
Since the original formulation of the Ho\v{r}ava theory in Ref.~\cite{Horava:2009uw}, the anisotropic conformal symmetry introduced in that paper has played an interesting and intriguing role (in $2+1$ dimensions these transformations were introduced in Ref.~\cite{Horava:2008ih}). The Weyl transformations are anisotropic in the sense that the lapse function scales with a weight different from the one of the spatial metric and the shift vector,
\begin{equation}
\tilde{g}_{ij} = \Omega^2 g_{ij} \,,
\hspace{2em}
\tilde{N} = \Omega^{3} N \,,
\hspace{2em}
\tilde{N}_i = \Omega^2 N_i \,,
\label{introweyl}
\end{equation}
where $\Omega = \Omega(t,\vec{x})$. This is related to the fact that the whole Ho\v{r}ava theory is anisotropic between space and time (see discussion in Ref.~\cite{Griffin:2011xs}), which is the core for its (power-counting) renormalizability. When introducing the anisotropic Weyl transformations, Ho\v{r}ava noticed that the coupling constant of the kinetic term of the Lagrangian,
 \begin{equation}
  \sqrt{g} N ( K_{ij} K^{ij} - \lambda K^2 ) \,,
 \end{equation}
must be fixed to $\lambda = 1/3$ (in 3 spatial dimensions), in order for this kinetic term to be effectively conformal. This gives a special role to this value of $\lambda$. To have a complete conformal theory the potential, of course, must be chosen to be conformal.

The anisotropic conformal Ho\v{r}ava theory, defined by two conditions: $\lambda =1/3$ and the setting of a conformal potential, has been broadly studied in several contexts, see, for example, \cite{Griffin:2011xs,Park:2009hg,Hartong:2015zia}. In particular, in Ref.~\cite{Griffin:2011xs} it arose in the context of holographic renormalization of relativistic gravity with asymptotically Lifshitz spacetimes: Anisotropic conformal terms emerge as counterterms. On the other hand, the case of the Ho\v{r}ava theory with $\lambda = 1/3$ fixed but \emph{without} a conformal potential has been studied independently; see, for example, \cite{Bellorin:2013zbp,Bellorin:2016wsl}. Since in this case only the kinetic part of the Lagrangian acquires the anisotropic conformal symmetry, it has been called the kinetic-conformal Ho\v{r}ava theory. We stress that the kinetic-conformal Ho\v{r}ava theory is not a conformal gravitational theory. Note that the anisotropic Weyl scalings (\ref{introweyl}), where the scale factor has an arbitrary dependence on the space, are defined for the nonprojectable version of the Ho\v{r}ava theory. The definition of the Ho\v{r}ava theory in the nonprojectable version was extended in Ref.~\cite{Blas:2009qj}.

An interesting feature of the kinetic-conformal theory is that it possesses the same number of degrees of freedom of General Relativity (GR). This was shown in Ref.~\cite{Bellorin:2013zbp} and corroborated with perturbative computations and an explicit $z=3$ potential in \cite{Bellorin:2016wsl}. Hence, the so-called extra mode of the Ho\v{r}ava theory, which has been the subject of intense study (see for example the issue of the strong coupling \cite{Charmousis:2009tc,Papazoglou:2009fj,Blas:2009ck}), is absent in the kinetic-conformal theory.  We comment that this is the motivation to regard the kinetic-conformal theory as a dynamically independent formulation of the Ho\v{r}ava theory, hence its different name, although it is just the nonprojectable Ho\v{r}ava theory at the critical point $\lambda=1/3$. The elimination of the extra mode is a purely dynamical effect. This means that the set of constraints of the theory, which are of first and second class, are responsible for the suppression. In particular, the kinetic-conformal theory has two additional second-class constraints that are absent in the generic ($\lambda \neq 1/3$) Ho\v{r}ava theory; they drop the extra mode. In general the second-class constraints are not, in principle, related to gauge symmetries.

However, it is worth studying the possible relationship between the anisotropic Weyl transformations, the anisotropic conformal Ho\v{r}ava theory, and its kinetic-conformal formulation. In particular, it is intriguing that one of the second-class constraints of the kinetic-conformal theory we have mentioned is precisely the generator of the Weyl scalings on the spatial metric. Of course, in the case of the kinetic-conformal theory this constraint must be of second class since the exact Weyl symmetry is broken. But it should also arise in the anisotropic conformal theory as a first-class constraint. Hence, at least part of the structure of the constraints is shared by these two versions of the Ho\v{r}ava theory.

Our aim in this paper is to take some steps towards the understanding of the effect of the anisotropic Weyl transformations in the dynamics of the Ho\v{r}ava theory, both in the anisotropic conformal and the kinetic-conformal formulations. Our central study is the canonical formulation of the anisotropic conformal theory. Despite the fact that a thorough analysis on the conformal anomaly for this theory is still missing,\footnote{The breakdown of the Weyl symmetry in the perturbative quantum theory was found in a specific anisotropic conformal Ho\v{r}ava model in Ref.~\cite{Adam:2009gq}.} we may start by performing a canonical analysis at the classical level. In particular, this will allow us to determine the number of degrees of freedom propagated by the conformal theory in the sense of initial data. For the sake of generality, we first develop an analysis with a general conformal potential. Next, we consider two models with specific conformal potentials. Part of our interest is in contrasting the structure of the set of constraints and the equations of motion of the anisotropic conformal Ho\v{r}ava theory with the kinetic-conformal case. As we have commented, there should be a first-class constraint playing the role of the generator of the anisotropic Weyl scalings. We study this constraint and analyze the similarities and differences between both theories when implementing the Dirac procedure of preserving the constraints in time. Throughout our analysis, we find interesting geometric structures associated to the anisotropic Weyl transformations.

Another way to explore the connection between the broken Weyl transformations and the dynamics of the kinetic-conformal theory is to study conformally flat solutions. We first point out that arbitrary anisotropic Weyl scalings on the Minkowski solution, that is,
\begin{equation}
 \tilde{g}_{ij} = \Omega^2 \delta_{ij} \,,
 \hspace{2em}
 \tilde{N} = \Omega^{3} \,,
 \hspace{2em}
 \tilde{N}_i = 0 \,,
 \label{confflatminkows}
\end{equation}
define a class of \emph{anisotropic} conformally flat configurations. It is clear that all configurations of this class are solutions of the anisotropic conformal theory due to its anisotropic Weyl symmetry. But the same does not need to be true in the kinetic-conformal theory since this symmetry is broken. On the other hand, there is no spacetime metric in the Ho\v{r}ava theory, the spatial metric is the natural geometric object instead. This suggests limiting the definition of conformal flatness to the spatial metric, leaving for the lapse function more possibilities than the one dictated by (\ref{confflatminkows}). This way of defining conformal flatness leads, in principle, to a broader set of configurations. We study general features of these configurations in the anisotropic conformal and the kinetic-conformal theories. In the last case, we look for explicit solutions belonging to this class. We find important similiarities between these two theories with regard to the features of these configurations.

This paper is organized as follows. In section \ref{sec:conformalgeneral} we perform the Hamiltonian formulation of the anisotropic conformal Ho\v{r}ava theory. We first analyze the general theory and, subsequently, two explicit models. In section \ref{sec:hamiltoniankc} we summarize the Hamiltonian formulation of the kinetic-conformal theory and contrast it with the dynamics of the anisotropic conformal theory. In section \ref{sec:confflat} we study the conformally flat solutions in both formulations of the Ho\v{r}ava theory.


\section{Hamiltonian analysis of the anisotropic conformal theory}
\label{sec:conformalgeneral}
\subsection{The anisotropic Weyl transformations}
\label{sec:conformal}
The fundamental assumption in Ho\v{r}ava's gravitational theory \cite{Horava:2009uw} is that there is an identified line of time, and a foliation of spatial hypersurfaces along the line of time is given. Since the line of time is identified, the underlying gauge symmetry of the theory consists of all the diffeomorphisms that preserve the foliation (FDiff). If a local coordinate system is used, this symmetry is defined by
\begin{equation}
\delta t = f(t) \,,
\hspace{2em}
\delta x^i = \zeta^i(t,\vec{x}) \,.
\end{equation}
As a consequence, in Ho\v{r}ava theory the spacetime pseudo-Riemannian metric characteristic of GR does not exist. Despite this fact, the theory is formulated in terms of the usual Arnowitt-Deser-Misner variables $N(t,\vec{x})$, $N_i(t,\vec{x})$ and $g_{ij}(t,\vec{x})$, but in this case $N$ and $N_i$ are interpreted as fields over the leaves of the foliation, and $g_{ij}$ is the Riemannian metric of the leaves of the foliation for each instant of time. We comment that the FDiff always allow us to choose a gauge in which $N_i = 0$, as in GR. Another consequence of this gauge symmetry is that the dependence in the space of the lapse function $N$ leads to two different formulations of the theory: the projectable and the nonprojectable versions. We study the nonprojectable case, where $N$ can depend on the spatial coordinates.

The core for the renormalizability of the theory, at least under power-counting analysis\footnote{The complete renormalization of the projectable Ho\v{r}ava theory has been shown in Ref.~\cite{Barvinsky:2015kil}.}, is the usage of a potential $\mathcal{V}$ containing only spatial derivatives \cite{Horava:2009uw}. For the power-counting renormalizability, in 3 spatial dimensions, terms of at least 6th order in spatial derivatives must be included in the potential ($z=3$ terms, in the nomenclature of \cite{Horava:2009uw}). Here, $\mathcal{V}$ must be a scalar under the FDiff, it can be written in terms of tensors over the spatial hypersurfaces that also depend on time. These tensors are $g_{ij}$, its covariant derivative and curvature tensors, and the FDiff-covariant vector 
\begin{equation}
 a_i = \frac{ \partial_i N }{N} \,,
 \label{ai}
\end{equation}
which plays a central role in the formulation of the nonprojectable theory \cite{Blas:2009qj}. The action of the complete nonprojectable theory is \cite{Horava:2009uw,Blas:2009qj}
\begin{equation}
S = \int dt d^3x \sqrt{g} N 
\left( G^{ijkl} K_{ij} K_{kl} - \mathcal{V} \right),
\label{lagrangianaction}
\end{equation}
where
\begin{eqnarray}
&& K_{ij} = \frac{1}{2N} ( \dot{g}_{ij} - 2 \nabla_{(i} N_{j)} ) \,,
\\[1ex]
&& G^{ijkl} = 
\frac{1}{2} \left( g^{ik} g^{jl} + g^{il} g^{jk} \right) 
- \lambda g^{ij} g^{kl} \,,
\end{eqnarray}
and $\lambda$ is a dimensionless constant. The dot stands for the time derivative, $\dot{g}_{ij} = \partial_t g_{ij}$.

In this paper we deal with the anisotropic conformal and the kinetic-conformal formulations of the Ho\v{r}ava theory. They share in common the condition
\begin{equation}
\lambda = 1/3
\end{equation}
(when the spatial dimension is $d$, the corresponding definition is $\lambda = 1/d$). When $\lambda$ takes this value the hypermatrix $G^{ijkl}$ becomes degenerated, that is,
\begin{equation}
 G^{ijkl} g_{kl} = 0 \,.
 \label{nullG}
\end{equation}
Furthermore, at this value the kinetic term of the Lagrangian, $\sqrt{g} N G^{ijkl} K_{ij} K_{kl}$, which is universal for all the Ho\v{r}ava models, gets an anisotropic conformal symmetry defined by the anisotropic Weyl scalings \cite{Horava:2009uw,Horava:2008ih}
\begin{equation}
\tilde{g}_{ij} = \Omega^2 g_{ij} \,,
\hspace{2em}
\tilde{N} = \Omega^{3} N \,,
\hspace{2em}
\tilde{N}_i = \Omega^2 N_i \,,
\label{conformaltrans}
\end{equation}
where $\Omega = \Omega(t,\vec{x})$. Then the conformal weights of $g_{ij}$ and $N_i$ are $+1$ and the one of $N$ is $+3/2$. If the condition $\lambda =1/3$ is set but the potential of the theory is not conformal, which is the case for a general FDiff-invariant potential, the theory is not Weyl invariant. This is the kinetic-conformal Ho\v{r}ava theory (or simply the nonprojectable Ho\v{r}ava theory at the critical point $\lambda = 1/3$). If, in addition, a conformal potential is chosen, the theory becomes anisotropic conformal. In other words, the kinetic-conformal theory is a theory with explicit breaking of the anisotropic Weyl symmetry. Note that, in order for the combination $\sqrt{g} N \mathcal{V}$ to be Weyl invariant, the potential $\mathcal{V}$ must have conformal weight $-3$, $\tilde{\mathcal{V}} = \Omega^{-6} \mathcal{V}$. One such conformal potential, which is also $z=3$, is \cite{Horava:2009uw}
\begin{equation}
 \mathcal{V} = \omega C^{ij} C_{ij} \,,
 \label{cottonintro}
\end{equation}
where $\omega$ is an arbitrary constant and $C^{ij}$ is the Cotton tensor
\begin{equation}
 C^{ij} = 
 \frac{1}{\sqrt{g}} \varepsilon^{kl(i} \nabla_k R_l{}^{j)} \,.
\end{equation}
The Cotton tensor has conformal weight $-5/2$, $\tilde{C}^{ij}=\Omega^{-5} C^{ij}$. We analyse in detail the model defined with this potential in section \ref{sec:cotton}.

Interestingly, the vector $a_i$ defined in (\ref{ai}) transforms as a gauge connection under the anisotropic Weyl scalings,
\begin{equation}
 \tilde{a}_i = a_i + 3 \frac{\partial_i \Omega}{\Omega} \,,
\end{equation}
as does the Levi-Civita connection of $g_{ij}$,
\begin{equation}
 \tilde{\Gamma}_{ij}^k = 
 \Gamma_{ij}^k 
 + \left( 2 \delta_{(i}^k \delta_{j)}^l - g_{ij} g^{kl} \right) \frac{\partial_l \Omega}{\Omega} \,.
\end{equation}
Hence in the anisotropic conformal Ho\v{r}ava theory one has two natural gauge connections for the conformal transformations. The composition of covariant conformal derivatives must be done in a way that is compatible with the symmetry of spatial diffeomorphisms. If $\Psi$ is a scalar under spatial diffeomorphisms and has conformal weight $r$,
\begin{equation}
 \tilde{\Psi} = \Omega^{2r} \Psi,
\end{equation}
we may define a conformal covariant derivative acting on $\Psi$ by
\begin{equation}
 D_i \Psi \equiv \partial_i \Psi - \frac{2r}{3} a_i \Psi \,.
 \label{confcovder}
\end{equation}
This derivative has conformal weight $r$ and it is also a vector under spatial diffeomorphisms. Here, $\Gamma_{ij}^k$ may also arise as a conformal gauge connection. For an arbitrary conformal spatial vector $\Psi_i$ of conformal weight $r$, under a Weyl transformation its covariant derivative with the Levi-Civita connection transforms as
\begin{equation}
 \tilde{\nabla}_i \tilde{\Psi}_j =
 \Omega^{2r} \nabla_i \Psi_j
 + \left[ ( 2r - 1 ) \delta_i^k \delta_j^l - \delta_j^k \delta_i^l
 + g_{ij} g^{kl} \right] \Omega^{2r-1} \partial_k \Omega \,\Psi_l \,.
\end{equation}
Now, for the specific case where the vector $\Psi_i$ has conformal weight $r = -1/2$ ($\tilde{\Psi}_i = \Omega^{-1} \Psi$), the hypermatrix in the second term of this expression becomes degenerated, with $g_{ij}$ being its null eigenvector,
\begin{equation}
 \left[ ( 2r - 1 ) \delta_i^k \delta_j^l - \delta_j^k \delta_i^l
+ g_{ij} g^{kl} \right] g^{ij} \equiv 0
 \quad \mbox{if} \quad 
 r= -1/2 \,.
\end{equation}
Therefore, the divergence of any spatial conformal vector of conformal weight $-1/2$,
\begin{equation}
 \nabla_k \Psi^k \,,
 \label{confdivergence}
\end{equation}
is also a conformal object with conformal weight $-3/2$, $\tilde{\nabla}_k \tilde{\Psi}^k = \Omega^{-3} \nabla_k \Psi^k$. In section \ref{sec:conformalmodel}, where we deal with an explicit conformal model, we see how the two conformal constructions (\ref{confcovder}) and (\ref{confdivergence}) arise explicitly. Moreover there is a third object transforming as the time component of a gauge connection for the anisotropic Weyl scalings, which we show it in section \ref{sec:canonicalgeneral}. 


\subsection{Canonical formulation}
\label{sec:canonicalgeneral}

We start the Hamiltonian formulation of the anisotropic conformal Ho\v{r}ava theory. The analysis is very close to that of the kinetic-conformal theory presented in Ref.~\cite{Bellorin:2013zbp}, but with important differences introduced by the additional Weyl symmetry. We set $\lambda = 1/3$ and assume that we are provided with a general conformal potential $\mathcal{V}$ that depends both on the metric $g_{ij}$ and the vector $a_i$ (unlike (\ref{cottonintro}), that does not depend on $N$). The canonically conjugated pairs are $(g_{ij},\pi^{ij})$ and $(N,P_N)$. Under the Weyl symmetry the momenta transform with the opposite weight of their conjugated variables. Since the Lagrangian does not depend on the time derivative of $N$, we have the primary constraint 
\begin{equation}
P_N = 0 \,.
\label{pncero}
\end{equation}
The canonical momentum conjugated to the spatial metric satisfies
\begin{equation}
\frac{\pi^{ij}}{\sqrt{g}} = 
G^{ijkl} K_{kl} \,.
\label{momentum}
\end{equation}
By using (\ref{nullG}), and denoting the trace as $\pi \equiv g_{ij} \pi^{ij}$, we get another primary constraint
\begin{equation}
\pi = 0 \,.
\label{picero}
\end{equation}
Although the two constraints (\ref{pncero}) and (\ref{picero}) are also present in the kinetic-conformal theory \cite{Bellorin:2013zbp}, the anisotropic Weyl symmetry modifies their behavior. Indeed, since $NP_N$ and $\pi$ are the generators of the Weyl scalings on $(N,P_N)$ and $(g_{ij},\pi^{ij})$ respectively, we have that in the anisotropic conformal theory the combination,
\begin{equation}
 \varpi \equiv \pi + \frac{3}{2} N P_N \,,
 \label{generatorweyl}
\end{equation}
is the generator of the anisotropic Weyl scalings (the numerical factor follows from the conformal weights defined in (\ref{conformaltrans})). Hence, this combination is a first-class constraint. In the kinetic-conformal theory there is no first-class behavior associated with $\pi$ or $P_N$ since there is no Weyl symmetry. In the anisotropic conformal theory it is convenient to replace the constraint $\pi = 0$ by the constraint $\varpi = 0$. This is similar to the way the generator of the purely spatial diffeomorphisms of the full theory is defined as a combination of two constraints, see (\ref{momentumconstraintconf}).

The Hamiltonian, with the primary constraints added, is
\begin{equation}
H = 
  H_0 
+ \int d^3x \left( N_i \mathcal{H}^i + \mu \varpi + \sigma P_N \right) \,,
\label{H}
\end{equation}
where
\begin{equation}
  H_0 = 
  \int d^3x \left(
  \frac{N}{\sqrt{g}} \pi^{ij} \pi_{ij} 
+ \sqrt{g} N \mathcal{V} \right) \,.
 \label{nonzerohamiltonian}
\end{equation}
$\mathcal{H}^i = 0$ is a primary constraint defined by
\begin{equation}
\mathcal{H}^i \equiv - 2 \nabla_j \pi^{ij} + P_N \partial^i N \,.
\label{momentumconstraintconf}
\end{equation}
As in GR, this is a first-class constraint, it is the generator of purely spatial diffeomorphisms. $N_i$, $\mu$ and $\sigma$ are Lagrange multipliers. Note that $H_0$ remains exactly invariant under the anisotropic Weyl scalings (\ref{conformaltrans}), as expected.

It is important to know the transformation of the Lagrange multipliers under the anisotropic conformal transformations. We pay special attention to the behavior of $\mu$, since it is the multiplier of the first-class constraint that generates the conformal transformations, hence a relation between the transformation of $\mu$ and the time derivative of the conformal factor is expected. We first comment that the momentum constraint is not conformal, but transforms with an additional term proportional to the constraint $\varpi$,
\begin{equation}
 \tilde{\mathcal{H}}^i =
 \Omega^{-2} \mathcal{H}^i
 + 2 \Omega^{-3} \partial^i \Omega \,\varpi 
\end{equation}
(this in turn is related to the fact that $\varpi$ is a scalar density). Next, if on the canonical action
\begin{equation}
 S = \int dt d^3x \left( 
     \pi^{ij} \dot{g}_{ij} + P_N \dot{N} 
   - \frac{N}{\sqrt{g}} \pi^{ij} \pi_{ij} 
   - \sqrt{g} N \mathcal{V}
   - N_i \mathcal{H}^i - \mu \varpi - \sigma P_N    
     \right) \,,
\end{equation}
we perform an anisotropic Weyl scaling, we get
\begin{equation}
\begin{array}{rcl}
 \tilde{S} &=& {\displaystyle
 \int dt d^3x \left[
   \pi^{ij} \dot{g}_{ij} + P_N \dot{N} 
 - \frac{N}{\sqrt{g}} \pi^{ij} \pi_{ij} 
 - \sqrt{g} N \mathcal{V}
 - N_i \mathcal{H}^i
\right. }
\\[2ex] && {\displaystyle \left.
 - \left( \tilde{\mu} - \frac{2\dot{\Omega}}{\Omega} 
      + 2 N_k \frac{\partial^k \Omega}{\Omega} \right) \varpi
 - \Omega^{-3} \tilde{\sigma} P_N \right] } \,.
\end{array}
\end{equation}
Therefore, we define the transformations of the Lagrange multipliers as
\begin{eqnarray}
 \tilde{\mu} &=& 
 \mu
 + \frac{2\dot{\Omega}}{\Omega} 
 - 2 N_k \frac{\partial^k \Omega}{\Omega} \,,
 \label{mutransf}
\\
 \tilde{\sigma} &=& \Omega^3 \sigma \,,
\end{eqnarray}
such that the canonical action remains invariant under anisotropic Weyl transformations in the whole phase space. As expected, the transformation of the Lagrange multiplier $\mu$ accounts for the time derivative of the conformal factor. Indeed, at least in the $N_i = 0$ gauge, $\mu$ transforms as the time component of a gauge connection for anisotropic conformal transformations.

Now we impose the time preservation of the primary constraints. Since $\mathcal{H}^i$ and $\varpi$ are generators of gauge symmetries, their preservation holds with no more conditions. This implies that in the anisotropic conformal theory there is a constraint less than in the kinetic-conformal theory, since in the latter the preservation of $\pi = 0$ leads to an additional constraint\footnote{If in the conformal theory one deals with $\pi=0$ instead of $\varpi=0$, the statement equivalent to the preservation of $\varpi=0$ without further conditions is that, due to the Weyl symmetry, the bracket of $\pi$ with the Hamiltonian is proportional to the one of $P_N$, $\{ \pi , H \} = - \frac{3}{2} N \{ P_N , H \} $ (imposing the primary constraints after computing the brackets). Therefore, the preservation of $\pi$ leads to the same condition imposed by the preservation of $P_N$.} \cite{Bellorin:2013zbp}.

The preservation of $P_N = 0$ leads to the secondary constraint
\begin{equation}
 \mathcal{H} \equiv 
  \frac{ \pi^{ij} \pi_{ij} }{ \sqrt{g}}
  + \mathcal{U} = 0 \,,
 \label{hamiltonianconstraintconf}
\end{equation}
where we have introduced the derivatives of the potential,\footnote{For notational simplicity here we have defined $\mathcal{U}$ and $\mathcal{W}^{ij}$ as densities, unlike Ref.~\cite{Bellorin:2016wsl}.}
\begin{equation} 
 \mathcal{U} \equiv 
 \frac{\delta}{\delta N} \int d^3y \sqrt{g} N \mathcal{V} \,,
 \quad
  {\mathcal{W}}^{ij} \equiv 
  \frac{1}{N} 
  \frac{\delta}{\delta g_{ij}} \int d^3y \sqrt{g} N \mathcal{V}
  \,.
 \label{derivativespotential}
\end{equation}
We call (\ref{hamiltonianconstraintconf}) the Hamiltonian constraint. It is interesting to see that the anisotropic Weyl symmetry relates the two derivatives $\mathcal{U}$ and $\mathcal{W}^{ij}$. This can be directly derived from the fact that the constraint $\varpi$ is the generator of this gauge symmetry and consequently its bracket with the Hamiltonian is automatically zero, as we have just commented. By making the computations explicit, we get
\begin{equation}
 0 = \{ \varpi , H \} = - N ( \mathcal{W} 
                        + \frac{3}{2} \mathcal{U} ) \,, 
\end{equation}
where $\mathcal{W}$ is the trace of $\mathcal{W}^{ij}$, $\mathcal{W} \equiv g_{ij} \mathcal{W}^{ij}$, and the equalities hold on the constrained phase space. Since in the anisotropic conformal theory the vanishing of this bracket is an identity, we get the following identity relating the two derivatives $\mathcal{U}$ and $\mathcal{W}^{ij}$:
\begin{equation}
 \mathcal{W} = - \frac{3}{2} \mathcal{U} \,.
 \label{wu}
\end{equation}
In section \ref{sec:conformalmodel} we will check this identity explicitly. It is also interesting that in the kinetic-conformal theory the same relation (\ref{wu}) arises, but, since the theory does not have the anisotropic Weyl symmetry, it is not an identity but a \emph{constraint} of the theory. We discuss this in section \ref{sec:hamiltoniankc}.

Finally, the time preservation of the secondary constraint $\mathcal{H}$ leads to an equation for the Lagrange multiplier $\sigma$, namely,
\begin{equation}
 \int d^3y \frac{\delta \mathcal{U}}{\delta N} \sigma =
 2 \int d^3y \frac{N \pi_{ij}}{\sqrt{g}}  
 \frac{\delta \mathcal{U}}{\delta g_{ij}}
 - \frac{2 N }{\sqrt{g}} \pi_{ij} \mathcal{W}^{ij}  
  \,,
 \label{sigmaeqconf}
\end{equation}
where the equality holds over the totally constrained phase space. In expressions like this, functional derivatives of functions are understood in an integrated form, for example
\begin{equation}
\int d^3y \sigma(y)
\frac{\delta \mathcal{U}(x)}{\delta N(y)}  =
\int d^3y \sigma(y)
\frac{\delta }{\delta N(y)} \int d^3z \delta^{(3)}(z-x) \mathcal{U}(z) \,.
\end{equation}
Dirac's procedure for the time preservation of constraints ends with Eq.~(\ref{sigmaeqconf}). Summarizing, we have that the anisotropic conformal theory possesses the constraints
\begin{eqnarray}
&& P_N = 0 \,,
\\
&& \mathcal{H}^i = - 2 \nabla_j \pi^{ij} + P_N \partial^i N =0\,,
\label{momentumconstfinal}
\\
&& \varpi = \pi + \frac{3}{2} N P_N =0\,,
\label{varpicero}
\\
&&
\mathcal{H} =
\frac{ \pi^{ij} \pi_{ij} }{ \sqrt{g}}
+ \mathcal{U} = 0 \,.
\label{hamiltonianconstfinal}
\end{eqnarray}
They sum up six equations. There are two gauge symmetries, the spatial diffeomorphisms and the Weyl scalings, that lead to four gauge degrees of freedom. In the phase space spanned by $(g_{ij},\pi^{ij})$ and $(N,P_N)$, we are left with four independent degrees of freedom once the constraints are solved and the gauge symmetries fixed. This corresponds to two physical, even, independent degrees of freedom, which is the same number as GR and the kinetic-conformal Ho\v{r}ava theory.

The equations of motion in the Hamiltonian formalism, after imposing the $N_i = 0$ gauge condition, are
\begin{eqnarray} 
\dot{N} - \frac{3}{2} \mu N &=& \sigma  \,,
\label{dotnconf} 
\\
\dot{g}_{ij} - \mu g_{ij} &=& 
\frac{2 N}{\sqrt{g}} \pi_{ij}   \,,
\label{dotgconf}
\\
\dot{\pi}^{ij} + \mu \pi^{ij} &=& 
-\frac{2 N}{\sqrt{g}} ( \pi^{ik} \pi_k{}^j 
-\frac{1}{4} g^{ij} \pi^{kl} \pi_{kl} )
- N \mathcal{W}^{ij}
\,.
\label{dotpiconf}
\end{eqnarray}
Taking into account that the transformation of the Lagrange multiplier $\mu$ balances the time derivative of the conformal factor (it is a gauge connection, Eq.~(\ref{mutransf})), we find that these equations are manifestly anisotropic-Weyl invariant.

\subsection{An explicit $z=3$ conformal model}
\label{sec:conformalmodel}
Here we study our first explicit model. The symmetric tensor,
\begin{equation}
X_{ij} \equiv R_{ij} - \frac{2}{9} a_i a_j + \frac{4}{3} \nabla_{(i} a_{j)} \,,
\end{equation}
transforms under the anisotropic conformal transformations (\ref{conformaltrans}) as
\begin{equation}
\begin{array}{rcl}
\tilde{X}_{ij} &=& {\displaystyle
X_{ij}
- 4 \Omega^{-1} ( a_{(i} \partial_{j)} 
       - \frac{1}{3} g_{ij} a_k \partial^k ) \Omega
- 12 \Omega^{-2} ( \partial_i \Omega \partial_j \Omega 
- \frac{1}{3} g_{ij} \partial_k \Omega \partial^k \Omega ) }
\\[2ex] && {\displaystyle 
+ 3 \Omega^{-1} ( \nabla_{ij} - \frac{1}{3} g_{ij} \nabla^2 ) \Omega }
\,,
\end{array}
\end{equation}
where $\nabla_{ij\cdots k} = \nabla_i \nabla_j \cdots \nabla_k$. This implies that its trace, 
\begin{equation}
 X \equiv R - \frac{2}{9} a_k a^k + \frac{4}{3} \nabla_k a^k \,,
\end{equation} 
is conformal with conformal weight $-1$, $\tilde{X} = \Omega^{-2} X$. Therefore the potential
\begin{equation}
 \mathcal{V} =
 \gamma {X}^3 \,,
 \label{confpotential}
\end{equation}
where $\gamma$ is an arbitrary coupling constant, has conformal weight $-3$. Hence this potential is suitable to define an anisotropic conformal Ho\v{r}ava theory of the class we studied in the previous section. This potential depends both on the spatial metric $g_{ij}$ and the $a_i$ vector. Interestingly, $\mathcal{V}$ is of $z=3$ order, the minimal order required for the power-counting renormalizability of the Ho\v{r}ava theory in $3$ spatial dimensions.

We specialize the formulas of the previous section to the case of the theory defined by the potential (\ref{confpotential}). The action and the primary Hamiltonian of the model are given by
\begin{eqnarray}
 && S = \int dt d^3x \sqrt{g} N 
 \left( G^{ijkl} K_{ij} K_{kl} - \gamma X^3 \right) \,,
 \\
 && H_0 = 
  \int d^3x \left(
  \frac{N}{\sqrt{g}} \pi^{ij} \pi_{ij} 
  + \gamma \sqrt{g} N X^3 \right) \,.
\end{eqnarray}
The direct computation of the  derivatives of the potential yields
\begin{eqnarray}
 \frac{\mathcal{U}}{\sqrt{g}} &=&
 \gamma X^3
 + \frac{4 \gamma}{N} \left[
    \nabla^2 + \frac{1}{3} \nabla_k (a^k \cdot ) \right] (NX^2)    
    \,,
 \label{uconfmodel}
 \\
 \frac{\mathcal{W}^{ij}}{\sqrt{g}} &=&
 \frac{\gamma}{2} g^{ij} X^3
 - \frac{3 \gamma}{N} \left[
   X^{ij} - \nabla^{ij} + g^{ij} \nabla^2 
 - \frac{4}{3} \nabla^{(i} ( a^{j)} \cdot )
 + \frac{2}{3} g^{ij} \nabla_k ( a^k \cdot ) \right] (NX^2)
  \,,
 \nonumber \\
 \label{wconfmodel}
\end{eqnarray}
where the dot inside the operators means $\nabla(a\cdot) b = \nabla(ab)$.

It is illustrative to discuss on explicit grounds the anisotropic conformal covariance of these two derivatives of the potential, which enter in the definition of the Hamiltonian constraint and the equations of motion, especially for the role of the vector $a_i$ as a gauge connection for the anisotropic conformal transformations. We first note that the combination $NX^2$ is a conformal spatial scalar with conformal weight $-1/2$. Therefore its conformal covariant derivative can be defined according to (\ref{confcovder}), that is, 
\begin{equation}
 D_i (NX^2) = \partial_i (NX^2) + \frac{1}{3} a_i NX^2 \,.
\end{equation}
In turn, this covariant derivative is a conformal spatial vector of conformal weight $-1/2$. According to the case discussed in Eq.~(\ref{confdivergence}), its divergence is also conformal. This divergence is exactly the combination arising between brackets in the right-hand side of Eq.~(\ref{uconfmodel}), $\nabla_k D^k (NX^2)$. Hence the second term in (\ref{uconfmodel}) has conformal weight $-3$, like the term $\gamma X^3$. This makes the $\mathcal{U}$ derivative conformally covariant in a manifest way, with conformal weight $-3/2$. Furthermore, we can write the two equations (\ref{uconfmodel} - \ref{wconfmodel}) in a form closer to the manifest conformal covariance,
\begin{eqnarray}
\frac{\mathcal{U}}{\sqrt{g}} &=&
 \gamma X^3
+ \frac{4 \gamma}{N} \nabla_k D^k ( N {X}^2 )
\,,
\label{uconfmanifest}
\\
\frac{\mathcal{W}^{ij}}{\sqrt{g}} &=&
 \frac{\gamma}{2} g^{ij} X^3
 - \frac{3 \gamma}{N} \left[
 X^{ij} - \nabla^{(i} D^{j)} 
 + g^{ij} \nabla_k D^k  
 - \nabla^{(i} ( a^{j)} \cdot )
 + \frac{1}{3} g^{ij} \nabla_k ( a^k \cdot ) \right] (NX^2)
 \,.
 \nonumber \\
 \label{wconfmanifest}
\end{eqnarray}
The transformation of the 2nd and 4th terms of the last equation,
\begin{eqnarray}
\nabla_{(i} D_{j)} (NX^2) &\rightarrow&
   \left[ \Omega^{-1} \nabla_{(i} D_{j)} 
   - \Omega^{-2} \left(
       \partial_{(i} \Omega D_{j)} 
     - g_{ij} \partial_k \Omega D^k \right) \right] (NX^2) \,,
\\
  \nabla_{(i} (a_{j)} NX^2 ) &\rightarrow&
  \left[\Omega^{-1} \nabla_{(i} (a_{j)} \cdot )
  - \Omega^{-2} \left( 3 \partial_{(i} \Omega a_{j)} 
      - g_{ij} \partial_k \Omega a^k \right) \right.
\nonumber \\ && \left.
  + 3 \nabla_{(i} \left( \Omega^{-2} \partial_{j)} \Omega \cdot \right)
  - 3 \Omega^{-3} \left( 2 \partial_i \Omega \partial_j \Omega 
     - g_{ij} \partial_k \Omega \partial^k \Omega \right) \right] (NX^2) \,,
\nonumber \\
\label{sometransf}
\end{eqnarray}
combined with the trace of (\ref{sometransf}), compensate the transformation of $X^{ij}$, such that the derivative $\mathcal{W}^{ij}$ becomes conformal with conformal weight $-5/2$.

In addition, by taking the trace of Eq.~(\ref{wconfmanifest}), one may check easily that the two derivatives of the potential (\ref{uconfmanifest} - \ref{wconfmanifest}) satify the identity (\ref{wu}).

The full set of constraints of the model is
\begin{eqnarray}
&& P_N = 0 \,,
\\
&& \mathcal{H}^i = - 2 \nabla_j \pi^{ij} + P_N \partial^i N =0\,,
\\
&& \varpi = \pi + \frac{3}{2} N P_N =0\,,
\\
&&
\mathcal{H} =
\frac{ \pi^{ij} \pi_{ij} }{ \sqrt{g}}
+ \gamma \sqrt{g} X^3
+ 4 \gamma \frac{\sqrt{g}}{N} \nabla_k D^k ( N {X}^2 ) = 0 \,.
\end{eqnarray}
Equation (\ref{sigmaeqconf}), which results from the time preservation of the Hamiltonian constraint, takes the form
\begin{equation}
\begin{array}{rcl}
 {\displaystyle
 {N}^{-1} \mathcal{D}_1 \hat{\sigma} } &=&
 {\displaystyle
 \frac{1}{4} \pi^{ij} a_i \nabla_j ( N {X}^2 )
 - \frac{3 N}{16 \sqrt{g}} \pi_{ij} \mathcal{W}^{ij}
 }
 \\[1ex] &&
 {\displaystyle
 + \frac{1}{2} \mathcal{D}_2^{ij}
 \left( \frac{9}{8} N {X}^2 \pi_{ij}
 - {X} a^k \nabla_k ( N \pi_{ij} )
 + 3 {X} \nabla^2 ( N \pi_{ij} ) \right) \,, 
 }
 \end{array}
 \label{sigmaeqmodel}
\end{equation}
where
\begin{eqnarray}
\hat{\sigma} &\equiv& \frac{ \sqrt{g} }{N} \sigma \,,
\\
\mathcal{D}_1 &\equiv&
\nabla^2 ( N {X} \nabla^2 \cdot ) 
+ \frac{1}{3} \nabla_k ( a^k N {X} \nabla^2 \cdot )
- \frac{1}{3} \nabla^2 ( a^k N {X} \nabla_k \cdot )
+ \frac{1}{8} N \nabla_k ( {X}^2 \nabla^k \cdot )
\nonumber \\ &&
+ \frac{3}{4} \left( N {X}^2 \nabla^2
        + \nabla_k ( N {X}^2 ) \nabla^k \right)
- \frac{1}{9} \nabla_k ( a^k a^l N {X} \nabla_l \cdot ) \,,
\\ 
\mathcal{D}_2^{ij} &\equiv&
 \nabla^{ij} + \frac{4}{3} a^i \nabla^j - R^{ij} + \frac{2}{9} a^i a^j
 \,.
\end{eqnarray}
Equation (\ref{sigmaeqmodel}) leads to an elliptic partial differential equation for $\hat{\sigma}$. Indeed, if we divide it by $X$, the highest order derivative acting on $\hat{\sigma}$ results $\nabla^4$. Thus, the set of constraints is consistently closed. The equations of motion of the model are
\begin{eqnarray} 
\dot{N} - \frac{3}{2} \mu N &=& \sigma  \,,
\\
\dot{g}_{ij} - \mu g_{ij} &=& 
\frac{2 N}{\sqrt{g}} \pi_{ij}   \,,
\\
\dot{\pi}^{ij} + \mu \pi^{ij} &=& 
- \frac{2 N}{\sqrt{g}} ( \pi^{ik} \pi_k{}^j 
- \frac{1}{4} g^{ij} \pi^{kl} \pi_{kl} )
- \frac{\gamma}{2} \sqrt{g} g^{ij} N X^3
\nonumber \\ &&
+ 3 \gamma \sqrt{g} \left[
X^{ij} - \nabla^{(i} D^{j)} 
+ g^{ij} \nabla_k D^k  
- \nabla^{(i} ( a^{j)} \cdot )
+ \frac{1}{3} g^{ij} \nabla_k ( a^k \cdot ) \right] (NX^2)
\,.
\nonumber \\
\end{eqnarray}


\subsection{An odd $z=3$ conformal model}
\label{sec:cotton}
In this section we study the second explicit model, which exhibits a rather different behavior. The $z=3$ potential
\begin{equation}
	\mathcal{V} = \omega C^{ij} C_{ij} \,,
	\label{cotton2}
\end{equation}
where $\omega$ is an arbitrary constant and $C^{ij}$ is the Cotton tensor
\begin{equation}
	C^{ij} = 
	\frac{1}{\sqrt{g}} \varepsilon^{kl(i} \nabla_k R_l{}^{j)} \,,
\end{equation}
is another example of conformal potential of weight $-3$. It was considered by Ho\v{r}ava in his original formulation of the theory \cite{Horava:2009uw}, and since then, it has been the subject of intense study.

The action and the primary Hamiltonian are
\begin{eqnarray}
&& S = \int dt d^3x \sqrt{g} N 
\left( G^{ijkl} K_{ij} K_{kl} - \omega C^{ij} C_{ij} \right) \,,
\\
&& H_0 = 
\int d^3x \left(
\frac{N}{\sqrt{g}} \pi^{ij} \pi_{ij} 
+ \omega \sqrt{g} N C^{ij} C_{ij} \right) \,.
\end{eqnarray}
The derivatives of the potential take the form
\begin{eqnarray}
&& \mathcal{U} = 
 \omega \sqrt{g} C^{kl} C_{kl} = \sqrt{g} \mathcal{V} \,,
 \label{ucotton}
 \\
&& \mathcal{W}^{ij} =
 2 \omega \sqrt{g} ( C^{ik} C_k{}^j - \frac{1}{4} g^{ij} C^{kl} C_{kl} )
 + \frac{\omega}{N} \mathcal{D}_3{}^{(ij)kl} (N C_{kl}) \,,
\end{eqnarray}
where
\begin{equation}
\begin{array}{rcl}
 \mathcal{D}_3{}^{ijkl} &\equiv&
 - \varepsilon^{mnk} \left[
 g^{il} \nabla_m{}^j{}_n 
 - \frac{1}{2} g^{ij} ( \nabla_{mn}{}^l + \nabla_m{}^l{}_n )
 - g^{il} R^j{}_n \nabla_m
 + g^{il} \nabla_m R^j{}_n
 \right]
\\[1.5ex] && 
 + \varepsilon^{imk} \left[
	 \nabla_m{}^{jl} 
	 - g^{jl} \nabla^2 \nabla_m
	 + \nabla^l ( R^j{}_m \cdot ) 
	 - g^{jl} \nabla_n ( R^n{}_m \cdot )
	 \right]
 \,,
\end{array}
\end{equation}
and, for later use,
\begin{equation}
\begin{array}{rcl}
\mathcal{D}_3^\dagger{}^{ijkl} &\equiv&
  \varepsilon^{mnk} \left[
g^{il} \nabla_n{}^j{}_m 
- \frac{1}{2} g^{ij} ( \nabla^l_{nm} + \nabla_n{}^l{}_m )
- g^{il} R^j{}_n \nabla_m 
- 2 g^{il} \nabla_m R^j{}_n
\right]
\\[1.5ex] &&
	- \varepsilon^{imk} \left[
	\nabla^{lj}{}_m 
	- g^{jl} \nabla_m \nabla^2
	+ R^j{}_m \nabla^l  
	- g^{jl} R^n{}_m \nabla_n 
	\right]
 \,.
\end{array}
\end{equation}

In the general analysis shown in section \ref{sec:conformal}, the time preservation of the constraint $\mathcal{H}$, given in Eq.~(\ref{hamiltonianconstraintconf}), leads to an equation for the Lagrange multiplier $\sigma$, which is Eq.~(\ref{sigmaeqconf}). In this model the potential (\ref{cotton2}) has no dependence on the lapse function $N$, a situation that we did not consider in the general case. This implies that the Hamiltonian constraint does not depend on $N$. As a consequence, its time preservation generates no equation for Lagrange multipliers, but another constraint instead, which for the model we have at hand is given by
\begin{equation}
  \mathcal{G} \equiv 
    \pi_{ij} \mathcal{D}_3{}^{ijkl} ( N C_{kl} )
  - C_{kl} \mathcal{D}_3^\dagger{}^{ijkl} ( N \pi_{ij} )
  = 0 \,.
  \label{oddconstraint}
\end{equation}
Finally, since the constraint $\mathcal{G}$ depends on $N$, its time preservation leads to an equation for the Lagrange multiplier $\sigma$. This equation is
\begin{equation}
 \pi_{ij} \mathcal{D}_3{}^{ijkl} ( N C_{kl} \sigma )
 - C_{kl} \mathcal{D}_3^\dagger{}^{ijkl} ( N \pi_{ij} \sigma )
 = - \{ \mathcal{G} , H_0 \} \,.
\end{equation}
Dirac's procedure stops at this step. The full set of constraints of this model is
\begin{eqnarray}
&& P_N = 0 \,,
\\
&& \mathcal{H}^i = - 2 \nabla_j \pi^{ij} + P_N \partial^i N =0\,,
\\
&& \varpi = \pi + \frac{3}{2} N P_N =0\,,
\\
&&
\mathcal{H} =
\frac{ \pi^{ij} \pi_{ij} }{ \sqrt{g}}
+ \omega \sqrt{g} C^{kl} C_{kl} = 0 \,,
\\
&&
\mathcal{G} = 
\pi_{ij} \mathcal{D}_3{}^{ijkl} ( N C_{kl} )
- C_{kl} \mathcal{D}_3^\dagger{}^{ijkl} ( N \pi_{ij} ) 
= 0 \,.
\end{eqnarray}

In this particular model, we end up with the same four constraints $\mathcal{H}^i$, $P_N$, $\varpi$ and $\mathcal{H}$ that the general anisotropic conformal theory has, plus the additional constraint $\mathcal{G}$. They sum up for a total of seven equations. There are four gauge degrees of freedom. This leaves three physical degrees of freedom in the phase space, which means two physical propagating modes: one even and the other one odd. As we have discussed, the origin of the odd nature of one of the propagating modes lies in the choice of a conformal potential that does not depend on the lapse function $N$.

The equations of motion of the odd model take the form
\begin{eqnarray} 
\dot{N} - \frac{3}{2} \mu N &=& \sigma  \,,
\\
\dot{g}_{ij} - \mu g_{ij} &=& 
\frac{2 N}{\sqrt{g}} \pi_{ij}   \,,
\\
\dot{\pi}^{ij} + \mu \pi^{ij} &=& 
- \frac{2 N}{\sqrt{g}} ( \pi^{ik} \pi_k{}^j 
- \frac{1}{4} g^{ij} \pi^{kl} \pi_{kl} )
- 2 \omega \sqrt{g} N ( C^{ik} C_k{}^j - \frac{1}{4} g^{ij} C^{kl} C_{kl} )
\nonumber \\ &&
- \omega \mathcal{D}_3{}^{(ij)kl} (N C_{kl})
\,.
\end{eqnarray}

\section{Hamiltonian formulation of the kinetic-conformal theory}
\label{sec:hamiltoniankc}
In this section we present a summary of the Hamiltonian formulation of the kinetic-conformal theory, which was developed in Ref.~\cite{Bellorin:2013zbp}, and contrast it with the previous theory.

The bulk part\footnote{As in GR, boundary terms are needed for the differentiability of the Hamiltonian under general variations \cite{Bellorin:2013zbp}. Here we omit these terms.} of the Hamiltonian is
\begin{equation}
H = 
\int d^3x \left( \frac{N}{\sqrt{g}} \pi^{ij} \pi_{ij} 
+ \sqrt{g} N \mathcal{V} + N_i \mathcal{H}^i + \mu \pi + \sigma P_N
\right) \,.
\label{hamiltonianbulkfinal}
\end{equation}
$\mathcal{V}$ is now considered as a general \emph{nonconformal} potential  that depends on $g_{ij}$ and $N$ and that is of $z=3$ order. We present the full list of constraints in this theory, \footnote{The constraints $\mathcal{H}$ and $\mathcal{C}$ must be also added to the Hamiltonian (\ref{hamiltonianbulkfinal}), but putting the associated Lagrange multipliers equal to zero always leads to a subset of consistent solutions \cite{Bellorin:2017gzj}. The solutions we study in the following belong to this subset.}
\begin{eqnarray}
&& P_N = 0 \,,
\label{pn}
\\
&& \pi = 0 \,,
\label{piceroconstraint}
\\
&& \mathcal{H} =
\frac{\pi^{ij} \pi_{ij}}{\sqrt{g}}   
+ \mathcal{U} = 0 \,,
\label{hamiltonianconstraintgeneral}
\\
&& \mathcal{C} =
\frac{3\pi^{ij} \pi_{ij}}{2\sqrt{g}}  
- \mathcal{W} = 0 \,,
\label{cconstraint}
\\
&& \mathcal{H}^i = 
- 2 \nabla_j \pi^{ij} + P_N \partial^i N  = 0 \,,
\label{momentumconstraint}
\end{eqnarray}
where $\mathcal{U}$ and $\mathcal{W}^{ij}$ are defined by the same formulas (\ref{derivativespotential}) and $\mathcal{W}$ is the trace of $\mathcal{W}^{ij}$. The time preservation of the $\pi = 0$ constraint leads to the constraint $\mathcal{C}= 0$. The time preservation of $\mathcal{H}$ and $\mathcal{C}$ imposes two equations on the Lagrange multipliers $\mu$ and $\sigma$, namely
\begin{eqnarray}
\int d^3y \left(
\frac{\delta \mathcal{U}}{\delta N} \sigma
+ g_{ij} \frac{\delta \mathcal{U}}{\delta g_{ij}} \mu \right)
- \frac{3 \pi^{ij} \pi_{ij}}{2 \sqrt{g}} \mu
= 
- 2 \int d^3y \frac{N \pi_{ij}}{\sqrt{g}} \frac{\delta \mathcal{U}}{\delta g_{ij}}
+ 2 \pi^{ij} \mathcal{W}_{ij} \,, && 
\label{sigmaeq}
\\
\int d^3y \left( 
\frac{\delta \mathcal{W}}{\delta N} \sigma
+ g_{ij} \frac{\delta \mathcal{W}}{\delta g_{ij}} \mu \right)
+ \frac{9 \pi^{ij} \pi_{ij}}{4 \sqrt{g}} \mu  
= 
- 2 \int d^3y \frac{\pi_{ij}}{\sqrt{g}} \frac{\delta \mathcal{W}}{\delta g_{ij}} 
- 3 \pi^{ij} \mathcal{W}_{ij} \,. && 
\label{mueq}
\end{eqnarray}
The anisotropic conformal and the kinetic-conformal theory share  exactly the same momentum constraint $\mathcal{H}^i$ since this is the generator of purely spatial diffeomorphisms, a symmetry present in both theories (and in GR). This is the only first-class constraint both theories have in common. Indeed, they also share the constraints $P_N$ and $\pi$, but in the kinetic-conformal theory the combination $\varpi = \pi + {\frac{3}{2} N P_N}$ lacks its first-class behavior due to the explicit breaking of the anisotropic Weyl symmetry. The Hamiltonian constraint $\mathcal{H}$ is also present as a second-class constraint, but in the kinetic-conformal case it is not conformal. Finally, in the kinetic-conformal theory we have the additional second-class constraint $\mathcal{C}$. Note that the two constraints $\mathcal{H}=0$ and $\mathcal{C}=0$ imply the condition
\begin{equation}
\mathcal{W} = - \frac{3}{2} \mathcal{U} \,.
\end{equation}
This is the same relation of (\ref{wu}) of the anisotropic conformal theory, but the way this relation was derived here identifies it as a constraint, rather than an identity.

There are seven equations defined by the constraints (\ref{pn} - \ref{momentumconstraint}), and three gauge degrees of freedom corresponding to the spatial diffeomorphisms. This leaves four physical degrees of freedom in the phase space, corresponding to two physical even modes, as we commented in section \ref{sec:conformalmodel}. The fact that the number of physical degrees of freedom is preserved between the anisotropic conformal and the kinetic-conformal theories is merely due to the exchange of a second-class constraint in the latter ($\mathcal{C}=0$) by a gauge symmetry in the former (the anisotropic Weyl symmetry).

Finally, the equations of motion in the Hamiltonian formalism, after imposing the $N_i = 0$ gauge condition, are
\begin{eqnarray} 
\dot{N} &=& \sigma  \,,
\label{dotn} 
\\
\dot{g}_{ij} &=& 
\frac{2 N}{\sqrt{g}} \pi_{ij} + \mu g_{ij} \,,
\label{dotg}
\\
\dot{\pi}^{ij} &=& 
-\frac{2 N}{\sqrt{g}} ( \pi^{ik} \pi_k{}^j 
-\frac{1}{4} g^{ij} \pi^{kl} \pi_{kl} )
- \mu \pi^{ij}
- N \mathcal{W}^{ij}
\,.
\label{dotpi}
\end{eqnarray}
Although these equations are not conformal, they share the same structure of the equations of motion of the anisotropic conformal case. The difference is, of course, the choice of the potential that determines the derivative $\mathcal{W}^{ij}$. This only affects the Eq.~(\ref{dotpi}) (the change in the Eq.~(\ref{dotn}) is only apparent, one can always redefine $\sigma$).


\section{Conformally flat solutions}
\label{sec:confflat}
In this section we study the issue of the conformally flat solutions in the context of the Hamiltonian formulation of the anisotropic conformal and kinetic-conformal Ho\v{r}ava theories.

\subsection{On the Conformal flatness in Ho\v{r}ava theory}
An essential characteristic of the Ho\v{r}ava theory is that a spacetime metric is not needed as a fundamental object. Instead, the spatial metric has independent existence. Indeed, the anisotropic nature of the Weyl transformations is related to the anisotropy between space and time \cite{Griffin:2011xs}. Nevertheless, Minkowski spacetime, defined in Ho\v{r}ava theory by $N = 1$, $N_i = 0$ and $g_{ij} = \delta_{ij}$ (in Cartesian coordinates), is a vacuum solution. Thus, in this framework, two different kinds of configurations arise, which we call conformally flat. The first one is the anisotropic conformal class of the Minkowski solution, characterized by $N = \Omega^3$, $N_i = 0$ and $g_{ij} = \Omega^2 \delta_{ij}$. All configurations of this class are automatically solutions of the anisotropic conformal theory. The second kind arises when the condition of conformal flatness is limited to the spatial metric $g_{ij}$, which is the natural metric in Ho\v{r}ava theory. We may define the associated conformally flat class by the single condition
\begin{equation}
g_{ij} = \Omega^2 \delta_{ij} \,, \quad \Omega = \Omega(t,\vec{x}) \,,
\label{conformallyflat}
\end{equation}
and allow for a lapse function in principle different to $N = \Omega^3$. The anisotropic conformal class of the Minkowski solution does not need to be solution of the kinetic-conformal theory due to the explicit breaking of the anisotropic conformal symmetry. On the contrary, the conformally flat class defined by the weaker condition (\ref{conformallyflat}) leads to a greater set of configurations that deserves to be studied independently.

Here we study these configurations. We start with the anisotropic conformal theory, leaving the potential undetermined. This allows us to extract some general properties of these configurations which can be compared with the kinetic-conformal theory. Our first result about the conformally flat configurations defined by (\ref{conformallyflat}) can be deduced by combining Eqs.~(\ref{varpicero}) and (\ref{dotgconf}), using $P_N=0$ explicitly. By inserting (\ref{conformallyflat}) in (\ref{dotgconf}) we get the equation
\begin{equation}
\left( \frac{2 \dot{\Omega}}{\Omega} - \mu \right) \delta_{ij} =
\frac{2 N}{\Omega^3} \pi_{ij} \,.
\label{dotgconfflat}
\end{equation}
Now the constraint $\pi = 0$ (\ref{varpicero}) implies that the trace of this equation is zero, so we have
\begin{equation}
\mu = \frac{2 \dot{\Omega}}{\Omega} \,.
\label{mudotomega}
\end{equation} 
By putting this relation back in Eq.~(\ref{dotgconfflat}) we get the strong result that the canonical momentum vanishes for all the conformally flat solutions ($N$ is nonzero by assumption),
\begin{equation}
\pi^{ij} = 0 \,.
\end{equation}

Now we extract further consequences of this result. By dropping the canonical momentum from Eq.~(\ref{dotpiconf}) we get the condition $\mathcal{W}^{ij} = 0$, that is,
\begin{equation}
\frac{\delta}{\delta g_{ij}} \int d^3y \sqrt{g} N \mathcal{V} = 0 \,,
\label{deltagcero}
\end{equation}
whereas from the Hamiltonian constraint (\ref{hamiltonianconstfinal}) we get $\mathcal{U} = 0$, hence
\begin{equation}
\frac{\delta}{\delta N} \int d^3y \sqrt{g} N \mathcal{V} = 0 \,.
\label{deltancero}
\end{equation}
Equations (\ref{deltagcero} - \ref{deltancero}) state that the conformally flat solutions defined by (\ref{conformallyflat}) are critical points of the potential of the anisotropic conformal theory. Note that, since $\pi^{ij} = P_N = 0$, the constraints (\ref{momentumconstfinal}) and (\ref{varpicero}) are automatically solved. The equation of motion (\ref{dotnconf}), after use of Eq.~(\ref{mudotomega}), can be brought to the form
\begin{equation}
 \Omega^3 \partial_t ( \Omega^{-3} N ) = \sigma \,.
 \label{sigmaconfflat}
\end{equation} 
This and the equation of motion (\ref{mudotomega}) fix the Lagrange multipliers $\sigma$ and $\mu$ in terms of the time derivative of the lapse function and the conformal factor. We comment that for the anisotropic conformally flat solutions belonging to the class of the Minkowski spacetime one has $N=\Omega^3$, hence the Eq.~(\ref{sigmaconfflat}) yields $\sigma = 0$ for this class. With regards the conformally flat configurations defined by (\ref{conformallyflat}), the remaining equations to be solved by them are the criticality conditions (\ref{deltagcero}) and (\ref{deltancero}). These equations must be solved for the variables $\Omega$ and $N$.

\subsection{Conformally flat solutions in the general kinetic-conformal theory}
Now we study the conformally flat configurations defined by (\ref{conformallyflat}) in the kinetic-conformal theory.

Since the Eq.~(\ref{dotg}) and the constraint (\ref{piceroconstraint}) remain the same in both theories, we obtain the same two results of the anisotropic conformal theory:
\begin{equation}
 \mu = \frac{2 \dot{\Omega}}{\Omega} \,,
 \label{mudotomegakc}
\end{equation} 
and 
\begin{equation}
 \pi^{ij} = 0 \,.
\end{equation}
Again, Eqs.~(\ref{dotn}) and (\ref{mudotomegakc}) fix the Lagrange multipliers in terms of time derivatives of $N$ and $\Omega$. Following the same lines of the anisotropic conformal theory, we find that the Hamiltonian constraint (\ref{hamiltonianconstraintgeneral}) and the equation of motion 
(\ref{dotpi}) imply 
\begin{equation}
 \mathcal{U} = \mathcal{W}^{ij} = 0
 \label{criticalpointkc}
\end{equation}
for these configurations. Therefore, we halso find that in the kinetic-conformal theory the conformally flat configurations (\ref{conformallyflat}) are critical points of the potential. Constraint $\mathcal{C}$, given in (\ref{cconstraint}), implies $\mathcal{W}=0$, but this is already implied by $\mathcal{W}^{ij}=0$, therefore constraint $\mathcal{C}$ imposes no further conditions on the conformally flat configurations. The constraints (\ref{piceroconstraint}) and (\ref{momentumconstraint}) are again automatically solved. As in the case of the anisotropic conformal theory, the remaining conditions are the conditions (\ref{criticalpointkc}) of criticality of the potential on the variables $\Omega$ and $N$.


\subsection{The effective $z=1$ kinetic-conformal theory}
To further advance with the conformally flat configurations in the kinetic-conformal theory, here we take a specific model, which is the effective action for large distances. The dominant terms in the potential at large distances are the $z=1$ terms. The corresponding potential is
\begin{equation}
 \mathcal{V} = - \beta R - \alpha  a_i a^i \,,
\end{equation}
where $\beta,\alpha$ are constants. For this effective theory the derivatives of the potential take the form
\begin{eqnarray}
\frac{\mathcal{U}}{\sqrt{g}} &=&
- \beta R - \alpha a_k a^k + 2 \alpha \frac{\nabla^2 N}{N} 
\,,
\label{uz1}
\\
\frac{{\mathcal{W}}^{ij}}{\sqrt{g}} &=& 
- \frac{\beta}{N} ( \nabla^{ij} - g^{ij} \nabla^2 ) N
+ \beta ( R^{ij} - \frac{1}{2} g^{ij} R )
+ \alpha ( a^i a^j - \frac{1}{2} g^{ij} a_k a^k )  \,,
\label{wijz1}
\\
\frac{\mathcal{W}}{\sqrt{g}} &=&
2\beta \frac{\nabla^2 N}{N} 
- \frac{1}{2} ( \beta R + \alpha a_k a^k ) \,.
\label{wz1}
\end{eqnarray}

Now we specialize to the conformally flat solutions. The fields $\Omega$ and $N$ must satisfy the criticality conditions (\ref{criticalpointkc}). From Eqs.~(\ref{uz1}) and (\ref{wz1}) we get that conditions $\mathcal{U}=0$ and $\mathcal{W}= 0$ are equivalent to the two equations
\begin{eqnarray}
\nabla^2 N &=& 0 \,,
\label{harmonicN}
\\
\beta R + \alpha a_k a^k &=& 0 \,.
\label{tracew}
\end{eqnarray}
Now we may use these two equations to simplify the equation $\mathcal{W}^{ij} = 0$, with $\mathcal{W}^{ij}$ given in (\ref{wijz1}), such that we obtain the equation
\begin{equation}
 \beta \nabla_{ij} N - \beta N R_{ij} - \alpha N a_i a_j  = 0 \,.
 \label{finaleqwij}
\end{equation}
Thus, we see that the criticality conditions (\ref{criticalpointkc}) are equivalent to the Eq.~(\ref{finaleqwij}) and either Eq.~(\ref{harmonicN}) or (\ref{tracew}). We choose Eq.~(\ref{harmonicN}), such that Eq.~(\ref{tracew}) is implied by Eqs.~(\ref{harmonicN}) and (\ref{finaleqwij}).

Evaluated on conformally flat metrics in Cartesian coordinates, $g_{ij} = \Omega^2 \delta_{ij}$, Eqs.~(\ref{harmonicN}) and (\ref{finaleqwij}) become, respectively,
\begin{eqnarray}
&& \partial_k \left( \Omega \partial_k N \right) = 0 \,,
\label{laplacianconformal}
\\
&& \beta \left[ 
  \partial_{ij} N 
- 2 \frac{\partial_{(i} \Omega \partial_{j)} N}{\Omega}
+ \frac{N \partial_{ij} \Omega}{\Omega} 
- 2 \frac{N \partial_i \Omega \partial_j \Omega}{\Omega^2}
+ \frac{\delta_{ij}}{\Omega} \left( \partial_k \Omega \partial_k N
     + N \partial_{kk} \Omega \right) \right]
\nonumber
\\ &&
- \alpha \frac{\partial_i N \partial_j N}{N} = 0 \,.
\label{ricciconformal}
\end{eqnarray}
We may give an explicit exact solution to the pair of Eqs.~(\ref{laplacianconformal} - \ref{ricciconformal}). Equation (\ref{laplacianconformal}) suggests the ansatz
\begin{equation}
 N = \Omega^m \,,
 \label{nomegam}
\end{equation}
where $m$ is an undetermined number. Under this ansatz Eq.~(\ref{laplacianconformal}) can be written as
\begin{equation}
 \left(\frac{m}{m+1}\right) \partial_{kk} \Omega^{m+1} = 0 \,,
\end{equation}
which implies, assuming $m\neq 0$, that the function $\Omega^{m+1}$ is a harmonic function with respect to the Euclidean flat Laplacian $\partial_{kk}$. With this identification Eq.~(\ref{laplacianconformal}) is completely solved. We allow for a general, parametric, time dependence of the chosen harmonic function (recalling that the time derivatives of $\Omega$ and $N$ determine the Lagrange multipliers $\mu$ and $\sigma$, Eqs.~(\ref{dotn}) and (\ref{mudotomega})). Next we substitute the ansatz (\ref{nomegam}) in Eq.~(\ref{ricciconformal}). This yields
\begin{equation}
 \Omega \partial_{ij} \Omega + n \partial_i \Omega \partial_j \Omega = 0 \,,
 \label{riccipre}
\end{equation}
where
\begin{equation}
 n \equiv \frac{ ( \beta - \alpha ) m^2 - 3 \beta m - 2 \beta }
      { \beta ( m + 1 ) } \,.
\end{equation}
Equation (\ref{riccipre}) can be brought to a simpler form,
\begin{equation}
 (n + 1)^{-1} \partial_{ij} \Omega^{n+1} = 0 \,.
\end{equation}
Assuming again that the coefficient is not zero, the general solution of this equation is a linear function of the spatial coordinates:
\begin{equation}
 \Omega^{n+1} = L \equiv \vec{k}(t) \cdot \vec{x} + b(t) \,,
\end{equation}
where $\vec{k}(t),b(t)$ are arbitrary functions of time. But in the previous analysis of the Eq.~(\ref{laplacianconformal}) we ended with the condition that $\Omega^{m+1}$ must be a harmonic function. Since $L$ is a harmonic function, these two conditions can be met simultaneously if $n = m$. This sets an equation for $m$ in terms of the coupling constants of the theory, namely,
\begin{equation}
 \alpha m^2 + 4 \beta m + 2 \beta = 0 \,.
 \label{eqm}
\end{equation}
If $\alpha = 0$ the only solution is $m= 1/2$. Otherwise the solutions are
\begin{equation}
 m_{\pm} = 
 - \frac{2 \beta}{\alpha} \pm \sqrt{ \frac{4\beta^2}{\alpha^2} - \frac{2\beta}{\alpha}} \,.
\end{equation}
Summarizing, the conformally flat solutions, in the gauge $N_i = 0$, are given by $\pi^{ij} = 0$, $g_{ij} = \Omega^2 \delta_{ij}$, $\Omega = L^{\frac{1}{m+1}}$, $N =  L^{\frac{m}{m+1}}$, $L$ is the general linear and time-dependent function $L = \vec{k}(t)\cdot\vec{x} + b(t)$, and the number $m$ is given by the roots of Eq.~(\ref{eqm}).


\section*{Conclusions}
We have performed the Hamiltonian formulation of the classical anisotropic conformal Ho\v{r}ava theory. Two conditions define this theory, namely, the fixed value of the coupling constant of the kinetic term and the presence of a conformal potential. We have contrasted the resulting dynamics with a related formulation, the kinetic-conformal theory, which has the same value of the coupling constant, but the potential is not conformal, hence the anisotropic Weyl symmetry is explicitly broken. We have found important similarities between the dynamics of both theories, on the basis of which we can state that the kinetic-conformal theory ``remembers" the anisotropic Weyl symmetry. We highlight the interesting point that the generator of anisotropic Weyl transformations is present in both theories, but it changes from first-class to second class in the passage from the anisotropic conformal to the kinetic-conformal theory. The number of the propagating degrees of freedom (at least in the classical-field sense) is the same in both theories. This is due to the exchange of the anisotropic Weyl symmetry by an extra second-class constraint in the kinetic-conformal case. We have found an important exception. This happens when the conformal potential does not depend on $N$, as is the case of the well known $(\mbox{Cotton})^2$ potential. In this case there emerges an additional second-class constraint, which leads to an odd physical mode (a mode that is odd in the phase space), plus an even mode. We have studied in detail two explicit models, one with the conformal potential depending on $N$ and the $(\mbox{Cotton})^2$ model. 

In the anisotropic conformal theory several objects play the role of gauge connections of the anisotropic Weyl transformations. These are  the vector $a_i = \partial_i \ln N$, the Levi-Civita connection of the spatial metric, and the Lagrange multiplier of the generator of the anisotropic Weyl transformations. The $a_i$ serves as gauge connection for general conformal objects, but the Levi-Civita connection only under specific combinations. The Lagrange multiplier acts as the time component of a gauge connection (in the $N_i = 0$ gauge), and we have identified it in the canonical equations of motion.

We have posed the question of whether nontrivial conformally flat solutions may arise in these theories. We have found that limiting the definition of conformal flatness only to the spatial metric gives more possibilities. Indeed, this kind of conformally flat configuration arises with similar properties in both theories. In particular, in both cases they are critical points of the potential and have vanishing canonical momentum. We have found an explicit conformally flat solution in the kinetic-conformal Ho\v{r}ava theory.

Since in Ho\v{r}ava theory the kinetic term and the potential of the Lagrangian are not connected by gauge symmetries, we have the interesting possibility of defining an ``intermediate" theory whose kinetic term is the same as in the anisotropic conformal theory but with a nonconformal potential. We have seen that the broken Weyl symmetry leaves several traces on the kinetic-conformal theory. It would be interesting to further explore this relationship. For example, many models of field theories with anisotropic Lifshitz scaling have been studied in holography. For a recent review on this topic see, for example, \cite{Taylor:2015glc}. On the basis of the similarities we have seen in the dynamics with the anisotropic conformal Ho\v{r}ava theory, we suggest that perhaps the kinetic-conformal Ho\v{r}ava theory could play an interesting role in the anisotropic holography.



\end{document}